\documentstyle[12pt,epsf]{article}

\newcommand{\insertplot}[1]{ \begin{center}\leavevmode\epsfysize=
17.9cm
\epsfbox{#1}\end{center}} 
\begin{document}
\parskip=0.3cm
\begin{titlepage}

\hfill \vbox{\hbox{DFPD 00/TH/50}\hbox{UNICAL-TH 00/8}
\hbox{November 2000}}

\vskip 0.3cm

\centerline{\bf ANALYTIC MODEL OF REGGE TRAJECTORIES $~^\diamond$}

\vskip 0.7cm

\centerline{R.~Fiore$^{a\dagger}$, L.L.~Jenkovszky$^{b\S}$, 
V.~Magas$^{b,c\ddagger}$, 
F.~Paccanoni$^{d\ast}$ and A.~Papa$^{a\dagger}$}

\vskip .3cm

\centerline{$^{a}$ \sl  Dipartimento di Fisica, Universit\`a della Calabria,}
\centerline{\sl Istituto Nazionale di Fisica Nucleare, Gruppo collegato di Cosenza}
\centerline{\sl I-87036 Arcavacata di Rende, Cosenza, Italy}
\vskip .3cm
\centerline{$^{b}$ \sl  Bogolyubov Institute for Theoretical Physics,}
\centerline{\sl National Academy of Sciences of the Ukraine}
\centerline{\sl 03143 Kiev, Ukraine}
\vskip .3cm
\centerline{$^{c}$ \sl Department of Physics, Bergen University} 
\centerline{\sl Allegaten 55, N-5007, Norway}
\vskip .3cm
\centerline{$^{d}$ \sl  Dipartimento di Fisica, Universit\`a di Padova,}
\centerline{\sl Istituto Nazionale di Fisica Nucleare, Sezione di Padova}
\centerline{\sl via F. Marzolo 8, I-35131 Padova, Italy}

\vskip 0.3cm

\begin{abstract}
A model for a Regge trajectory compatible with the threshold behavior 
required by unitarity and asymptotics in agreement with analyticity 
constraints is given in explicit form. The model is confronted in the
time-like region with widths and masses of the mesonic resonances and, 
in the space-like region, the $\rho$ trajectory is compared with
predictions derived from $\pi-N$ charge-exchange reaction. Breaking of 
the exchange degeneracy is studied in the model and its effect on both 
the masses and widths is determined. \linebreak
\end{abstract}

\vskip .3cm

\hrule

\vskip.1cm

\noindent
$^{\diamond}${\it Work supported by the Ministero italiano
dell'Universit\`a e della Ricerca Scientifica e Tecnologica and by the
INTAS.}
\vfill
$
\begin{array}{ll}
^{\dagger}\mbox{{\it e-mail address:}} &
   \mbox{FIORE,PAPA~@CS.INFN.IT} \\
^{\S}\mbox{{\it e-mail address:}} &
 \mbox{JENK~@GLUK.ORG} \\
^{\ddagger}\mbox{{\it e-mail address:}} &
   \mbox{VLADIMIR~@FI.UIB.NO} \\
^{\ast}\mbox{{\it e-mail address:}} &
   \mbox{PACCANONI~@PD.INFN.IT}
\end{array}
$

\vfill
\end{titlepage}
\eject
\textheight 210mm
\topmargin 2mm
\baselineskip=24pt

\section{Introduction}
Recently a lot of activity has been dedicated to various studies of 
the Regge trajectories~\cite{DDFT,FPS,FJMPP,BBG,AAS,TN,GN,KKL,DGMP,M,BP}. 
Some researchers deal with the 
improvement and extension of the existing fits to hadronic total cross 
sections by including, apart from the slopes and 
intercepts of degenerate or non-degenerate linear Regge trajectories, 
also the relevant Chew-Frautschi plots~\cite{GN,KKL,DGMP}. The most complete and 
updated Chew-Frautschi plots for 
various (linear) trajectories can be found in paper~\cite{AAS}. The 
essential role of the non-linearity was recognized and scrutinized in 
the papers~\cite{FPS,FJMPP,BBG}, where studies of the analytic and asymptotic 
properties of the trajectories were also performed. The connection of the
Regge trajectories with dynamical models, based on non-relativistic 
potentials, hadronic strings, dual models etc. was studied recently in 
Refs.~\cite{BBG,BP}.

     The basic problem in constructing Regge trajectories is 
to combine the nearly linear rise of the real part
in the whole range of observed resonances with  the presence of 
a sizable imaginary part. The combination of these properties in a 
single analytic function seems 
to be difficult,
if not impossible, unless the linear part stops rising at larger values
of the argument. This is true for the model we present below.     

     Besides the experimental evidence, the dominant idea about the 
continuous rising linear trajectories found strong support from the
narrow-resonance (e.g. Veneziano) dual models, later replaced by hadronic
strings. The ``narrow resonance width" was treated in those models
 as an approximation, where 
resonances' widths are considered negligible in comparison with the
spacing between the neighbors. The data do not support this idea, thus 
ruling out the concept of continuously rising linear trajectories.

Exchange degeneracy (EXD) of the Regge trajectories (and the residues) with 
opposite signatures is an approximate symmetry based both on observations 
and theoretical speculations (the quark model, $SU(3)$ symmetry, duality). It 
is attractive as it relates various objects (trajectories, spectra of 
particles, residues, cross sections) reducing thus the number of the free 
parameters. The extent to which this symmetry is violated has 
been (and is still) debated ever since this symmetry was discovered.
It depends on 
the considered reaction (trajectory)  and the required precision (at the cost 
of the number of the free parameters introduced). In our analysis we consider 
both cases - exact EXD as well as its violation. 

     There are two basic theoretical restrictions, which must be satisfied by 
Regge trajectories - the threshold and asymptotic behaviour.
The behaviour at the threshold is determined by unitarity and was studied in 
Refs.~\cite{BZ,GP,Taylor,O}.
On the other hand, only bounds exist on the asymptotic behaviour of the 
trajectories. Analyticity sets an upper bound~\cite{GP} and there are some
plausible arguments~\cite{DP,DAMA,T,KOB} in favour of a stronger 
upper bound. They suggest a square root asymptotic 
behaviour on the physical sheet, $\alpha(s) \sim (-s)^{1/2}$, for $|s|\to 
\infty$, implying, as we shall see,  that the real part of the trajectories 
is bounded by 
a constant, i.e. that resonances ``terminate" at some point.

In the present paper we construct an explicit model for light 
unflavored mesonic trajectories
satisfying the above properties, i.e. our model has correct threshold and 
asymptotic behaviour and fits the data on both the masses and widths of the
observed mesonic resonances. We did not include baryonic
resonances, whose widths seem to show a regularity, $Im\ \alpha(s)\sim 
Re\ \alpha(s)$,
not shared by mesonic resonances~\cite{FPS,DP}. 

We apply our analysis of the meson trajectories to both the positive (particles
spectra) and negative (scattering region) values of their arguments (we denote
it by $s$, although our results are crossing symmetric in the Mandelstam
variables $s, t, u$) by fitting the parameters of the $\rho-a_2$ exchange 
degenerate trajectory to $\pi-N$ charge-exchange reaction data as in~\cite{BARN}. 

Although various aspects of mesonic trajectories have been already 
studied and  clarified in several papers~\cite{FPS,DP,DAMA,T,KOB} we are not 
aware of any complete and coherent solution of the problem, satisfying 
the correct threshold behaviour and fitting the widths of the mesonic trajectories.

\section{The model}

The properties of the trajectories following from analyticity and unitarity 
have been discussed in Refs.~\cite{BZ,GP,Taylor,O,DP}. 
In the complex $s$-plane the 
trajectory $\alpha(s)$ has only right-hand branch lines apart from 
branch points at the intersection with another singularity in the 
$l-$plane. The requirement that there are no crossing poles and that 
$\alpha(s)$ is bounded for $s\to \infty$, leads to a dispersion relation 
of the form 
\begin{equation}
\alpha(s)=\alpha(0)+\frac{s}{\pi}\int_0^{\infty}\,ds'\frac{{\cal I}m\, 
\alpha(s')}{s' (s'-s)}
\label{a2}
\end{equation}
where ${\cal I}m\,\alpha(s)>0$ since the pole must lie in the upper half 
$l$-plane. In other words, no complex singularity of the partial wave 
exists for real $l$. The boundedness of $\alpha(s)$ for $s\to \infty$ 
follows from the assumption that the amplitude $A(s,t)$ is bounded by a 
polynomial in $t$ and $s$. This restriction is equivalent to the 
condition that the amplitude, in the Regge form, should have no 
essential singularity at infinity in the cut plane. The condition that 
${\cal R}e\,\alpha(s)$ is bounded by a constant, for $s\to\infty$, leads to
\begin{equation}
|\alpha(s)|<M s^q,\qquad \mbox{for}\;s\to\infty,
\label{b2}
\end{equation}
with $q<1$ and $M$ an arbitrary constant. Equation~(\ref{a2}) implies 
that ${\cal R}e\,\alpha(s)$ cannot be exactly linear in $s$ but it does not give any 
information on the deviation from linearity unless a specific model has 
been selected for the imaginary part ${\cal I}m\,\alpha(s)$.

Unitarity requires the threshold behaviour
\begin{equation}
{\cal I}m\,\alpha(s) \propto (s-s_0)^{{\cal R}e\,\alpha(s_0)+1/2}\ ,
\label{c2}
\end{equation}
as shown in Ref.~\cite{BZ,GP,O}. For example, the first right-hand 
branch point will be at $s_0=(2 m_{\pi})^2$ for the $\rho$ trajectory, 
and $s_0=(3 m_{\pi})^2$ for the $\omega$. To build an 
explicit model we must constrain the asymptotic behaviour of the 
trajectory and find a rule to determine the contribution of several 
thresholds to ${\cal I}m\,\alpha(s)$. We borrow both properties of the model 
from Ref.~\cite{KOB}. First, we assume
 the square-root asymptotic on the physical sheet
\begin{equation}
|\alpha(s)|\sim s^{\frac{1}{2}},\qquad \mbox{for}\;s\to\infty,
\label{d2}
\end{equation}
which satisfies the condition (\ref{b2}) and is realized in dual 
models~\cite{DAMA,T}. Secondly, we assume
 additivity of threshold contributions
\begin{equation}
\alpha(s)=\alpha(0)+\sum_n\alpha_n(s),
\label{e2}
\end{equation}
where $\alpha_n(s)$ has only one threshold branch point on the physical 
sheet. We require this hypothesis, which seems to be
plausible, in order to set up a completely soluble model; 
at least we are not aware 
of a reasonable alternative. Contrary to~\cite{KOB}, we consider in the 
following only thresholds determined from  pions and other mesonic 
resonances.

We start from a simple analytical model, where the imaginary part 
of the trajectory is chosen as a sum of the single threshold terms
\begin{equation}
{\cal I}m\,\alpha(s)=\sum_n\,c_n (s-s_n)^{1/2} \left( \frac{s-s_n}{s} 
\right)^{{\cal R}e\,\alpha(s_n)} \theta(s-s_n),
\label{f2}
\end{equation}
with the correct asymptotic and threshold behaviour. 
In Eq.~(\ref{f2}) all $c_n$'s are positive~\cite{GP,O}. It should be 
stressed that this choice is a crude approximation to the 
complexity of the true imaginary part. For example, the exponent
${\cal R}e\,\alpha(s_n)$ should be a continuous function of $s$.

From the dispersion relation for the trajectory, 
\begin{displaymath}
{\cal R}e\,\alpha(s)=\alpha(0)+\frac{s}{\pi}\,PV\int_0^{\infty} 
\,ds'\,\frac{{\cal I}m\,\alpha(s')}{s' (s'-s)},
\end{displaymath}
where $PV$ means the Cauchy Principal Value of the integral,
the real part can be easily calculated~\cite{DP,BAT}. Defining
$\lambda_n={\cal R}e\,\alpha(s_n)$ we get:
\begin{displaymath}
{\cal R}e\,\alpha(s)=\alpha(0)+\frac{s}{\sqrt{\pi}} \sum_n 
c_n\frac{\Gamma(\lambda_n 
+3/2)}{\Gamma(\lambda_n+2)\sqrt{s_n}}\,{}_2F_1\left(1,1/2;\lambda_n+2;
\frac{s}{s_n}\right)\theta(s_n-s)+
\end{displaymath}
\begin{equation}
+\frac{2}{\sqrt{\pi}}\sum_n c_n \frac{\Gamma(\lambda_n+3/2)}{\Gamma
(\lambda_n+1)}\sqrt{s_n}\;{}_2F_1\left(-\lambda_n,1;3/2;
\frac{s_n}{s}\right)\theta(s-s_n).
\label{g2}
\end{equation}
From Eq.~(\ref{g2}) we can get the slope
\begin{displaymath}
{\cal R}e\,\alpha '(s)=\frac{1}{\sqrt{\pi}} \sum_n c_n 
\frac{\Gamma(\lambda_n+3/2)}{\Gamma(\lambda_n+2)\sqrt{s_n}} 
{}_2F_1\left(2,1/2;\lambda_n+2;\frac{s}{s_n}\right)\theta(s_n-s)+
\end{displaymath}
\begin{equation}
+\frac{4}{3\sqrt{\pi}}\sum_n c_n 
\frac{\Gamma(\lambda_n+3/2)}{\Gamma(\lambda_n)}\frac{s_n^{3/2}}{s^2} 
{}_2F_1\left(1-\lambda_n,2;5/2;\frac{s_n}{s}\right)\theta(s-s_n),
\label{h2}
\end{equation}
since the derivatives of the $\theta$-functions in Eq.~(\ref{g2}) cancel. 
Equations~(\ref{f2}) and (\ref{h2}) determine the width of the 
resonances through the relation
\begin{equation}
\Gamma(M^2)= \frac{{\cal I}m\,\alpha(M^2)}{M {\cal R}e\,\alpha '(M^2)}
\label{i2}
\end{equation}
where $\Gamma$ is the width of the resonance and $M$ is its mass.

Suppose now that there is a threshold, higher than all resonances,
at the position $s_x$ and with coefficient $c_x$. This threshold will
contribute to the real part and to the slope, but not to the 
imaginary part in the resonance region. Its relevance to the real part 
of the trajectories and to the widths of the resonance will become 
clear in the following.  Fits of the unknown parameters will be 
presented and discussed in the next Section. We only notice that Eq.~(\ref{g2}) 
gives the complete real part of the trajectory, both in the time-like 
and in the space-like regions.

Of particular interest is the behaviour of the trajectory above $s_x$. 
From Eqs.~(\ref{g2}-\ref{i2}) we can see that
for $s\gg s_x$ the ${\cal R}e\,\alpha$ reaches a saturation point and stays 
nearly constant.

\begin{figure}
\insertplot{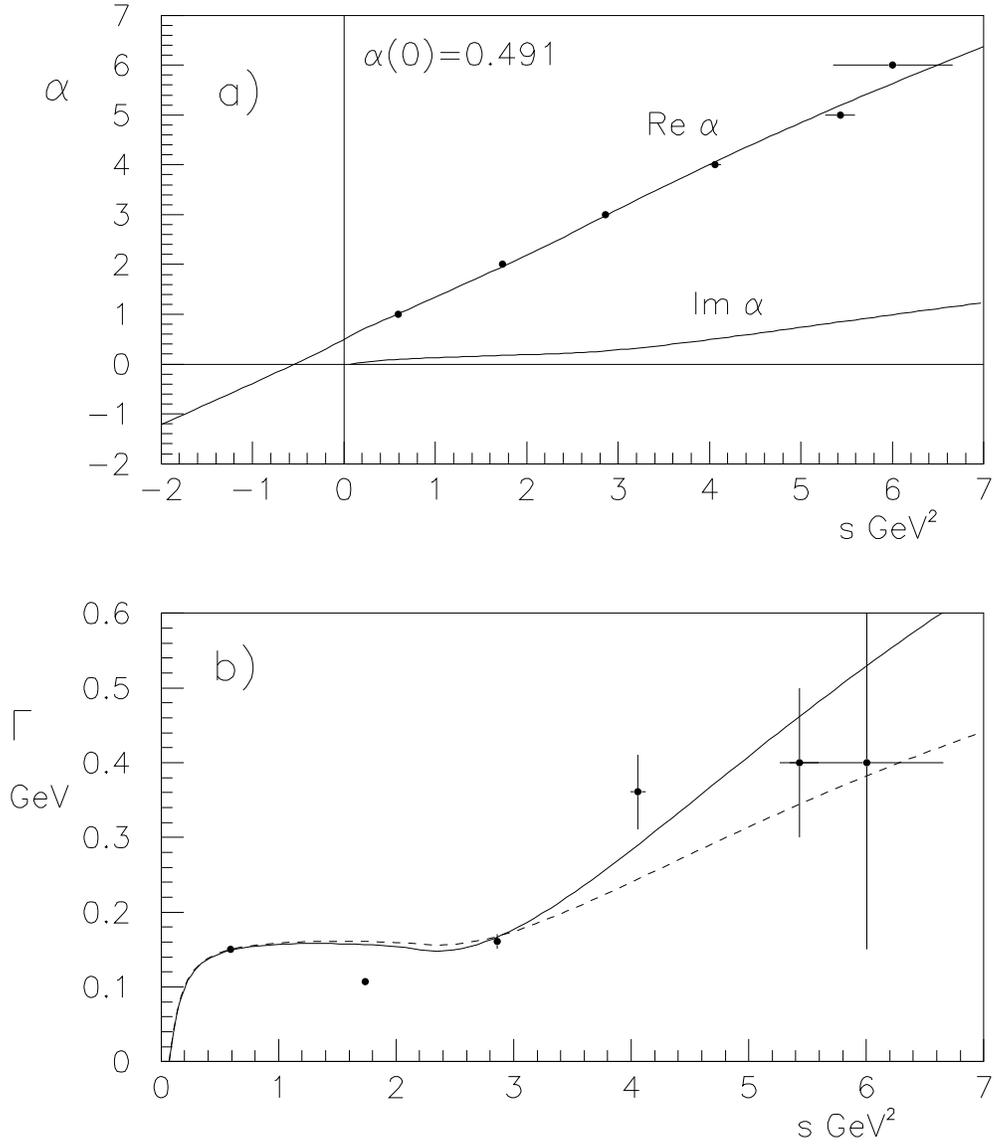}
\vspace{-1cm}
\caption{
(a)\ ${\cal R}e\ \alpha(s)$, 
${\cal I}m\ \alpha(s)$ for exchange degenerate 
$\rho-a_2$ trajectory; (b) continuous line:  $\Gamma(s)$ for the  
$\rho-a_2$ trajectory, dashed line: $\Gamma(s)$ for the  
non-degenerate $\rho$ trajectory. 
}
\label{fig1}
\end{figure}

\section{Fitting the resonances masses and widths}

The feature of the real part of trajectory, namely
 ${\cal R}e\,\alpha(s)$ tends to constant, when 
$s\to\infty$, means that 
the number of resonances lying on it is limited. As noticed before, 
approximate linearity in the region where experimental data require it, 
can be attained by introducing a threshold higher than the masses of all 
known relevant resonances. The highest threshold entails two unknown 
parameters, its position $s_x$ and the coefficient $c_x$, and 
contributes to Eq.~(\ref{f2}) only above the resonance region. 
It can be considered as an effective threshold. This fact 
suggests an iterative method for the fitting procedure.

A rough estimate of ${\cal R}e\,\alpha(s_n)\equiv \lambda_n$, can be 
obtained from a linear trajectory, adapted to the experimental data. 
In this way we fix also the lowest order value of the slope. Intercepts and 
slopes of mesonic trajectories can be taken, for example, from the 
accurate analysis in~\cite{DGMP}. In addition to the lowest 
$\pi-\pi$ threshold 
we will consider in the following the open channels, where one, or 
both pions, are substituted by mesonic resonances. Once the thresholds 
$s_n$ have been chosen, in accordance with the considered trajectory,
Eq.~(\ref{i2}) can be used to fit the unknown parameters 
$c_n$ from the experimental masses and widths of the resonances. Then, the 
fit of the 
spins of the resonances fix the remaining parameters, $\alpha(0)$, $c_x$ 
and $s_x$, from Eq.~(\ref{g2}). In the minimization procedure an 
error can be associated with the spin of a resonance in terms of the error 
in the mass and the estimated errors for intercept and 
slope~\cite{DGMP}. In such a way we get the first order correction to the 
trajectory. Iterations can be performed by repeating the previous steps 
till a stable output is obtained.

\begin{figure}
\insertplot{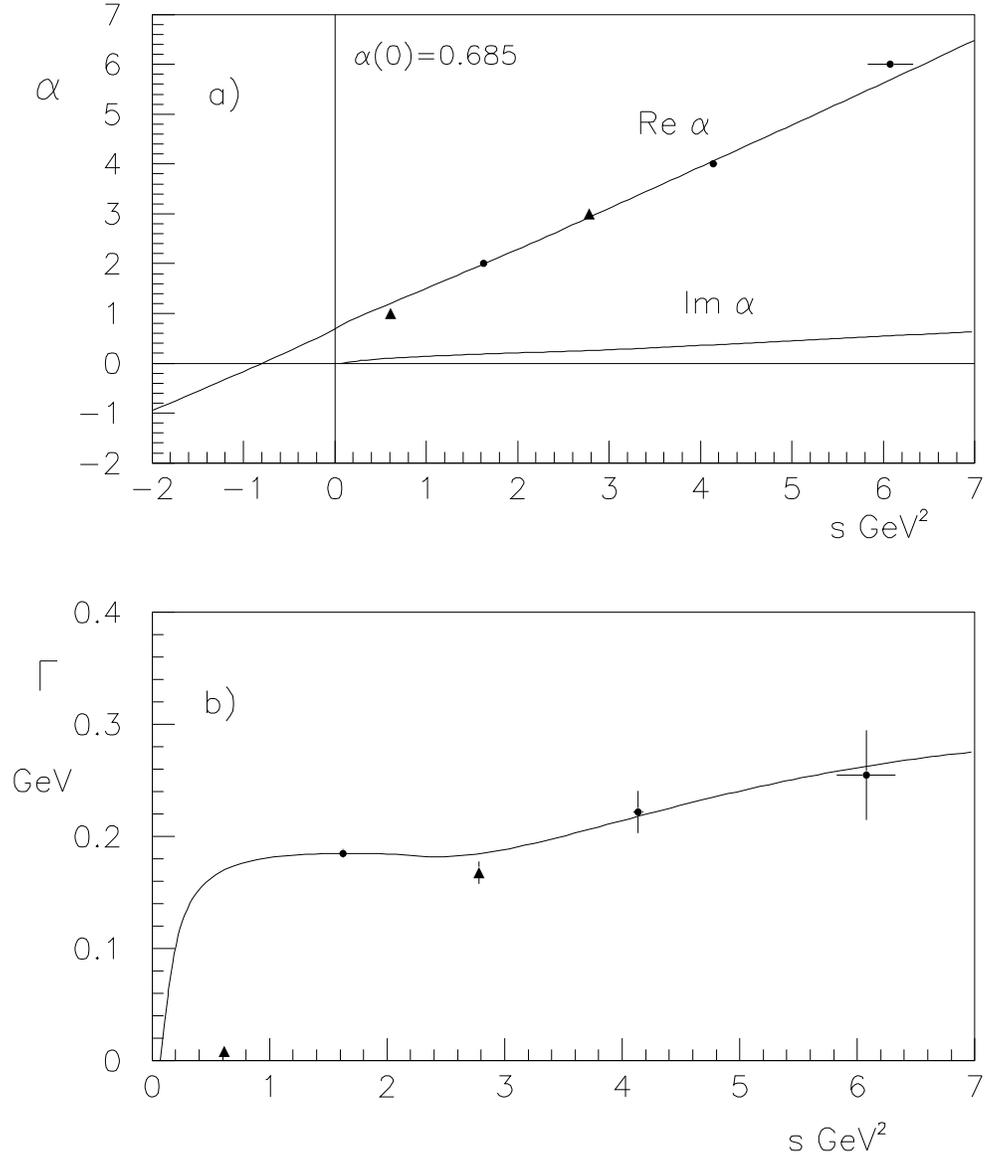}
\vspace{-1cm}
\caption{
(a)\ ${\cal R}e\ \alpha(s)$, 
${\cal I}m\ \alpha(s)$ for the (non-degenerate) 
$f$ trajectory; (b) $\Gamma(s)$ for the  
$f$ trajectory. In both figures triangles symbolize $\omega(782)$ and 
$\omega_3(1670)$, these resosonances have not been included in 
the fit.
}
\label{fig2}
\end{figure}

The scarcity of the available experimental data, 
forces an {\sl a priori} choice of the 
thresholds. In all cases with positive G-parity, the $\pi-\pi$ threshold 
will be present, hence $s_1\simeq 0.077$ GeV$^2$ for $\rho$ and $f$ 
trajectories. After a large number of trials we have chosen $s_2$ near 
 2 GeV$^2$ for all trajectories. To be precise, for the $G=+$ 
trajectories, we set $s_2=2.12$ GeV$^2$ that corresponds to a $\pi- 
a_2(1320)$ threshold. The position of the last threshold, $s_3\equiv 
s_x$ , does not influence sensibly the result. Hence, if the $\chi^2$ 
prefers a very high $s_x$ (we will see later that this is the case for 
$\rho$ and $\rho+a_2$ trajectories), we set $s_x=30$ GeV$^2$. We have tried to 
insert extra thresholds in a large variety of ways without finding any 
improvement. By fixing the thresholds we limit the capability of the 
model in reproducing the experimental data, but, as we will see in the 
following, this still allows us to obtain a good agreement 
with experiment.

Let us first consider the $\rho$ and $a_2$ trajectories. 
In fact only the EXD 
$\rho-a_2$ trajectory provides for a large set of input data: ten 
plus two, which need confirmation.
In our model $\rho$ and 
$a_2$ have different thresholds, with different G-parity, and exchange 
degeneracy is necessarily broken. Experimentally however the breaking is 
small and not easily detectable in the real part. As a first attempt, we 
keep all the resonances on an EXD  $\rho-a_2$ trajectory and fit them 
with five correction cycles. Data for the resonances $\rho(770),
\,a_2(1320),\,\rho_3(1690),\,a_4(2040),\,\rho_5(2350)$ and $a_6(2450)$ are 
taken from the Review of Particle Physics~\cite{CASO} and the zeroth order 
trajectory is written as ${\cal R}e\,\alpha_{\rho-a_2}(s_n) \equiv 
\lambda_n \approx 0.482+0.874\,s_n$. Notice that $a_6(2450)$ needs 
confirmation and $\rho_5(2350)$ has been omitted from the summary table.
Thresholds $s_1$ and $s_2$ are fixed as explained above, $s_x$ prefers 
to stay as large as possible and is fixed at 30 GeV$^2$.
In the output the intercept of the degenerate $\rho-a_2$ trajectory 
increased to $\alpha_{\rho-a_2}(0)\simeq 0.491$ while $c_1\simeq 0.140,\,c_2 
\simeq 0.902,\,c_x\simeq 28.031$. 
Figs.~\ref{fig1}(a, b) 
show, as continuous lines, ${\cal R}e\,\alpha(s)$, ${\cal I}m\,\alpha(s)$
 and $\Gamma(s)$ for the 
exchange degenerate $\rho-a_2$. In 
agreement with Eq.~(\ref{i2}), the widths of the resonances are affected
to a larger extent by a breaking of EXD. To substantiate this statement 
we performed a new fit, keeping only the $\rho$ resonances, with the same 
thresholds. While the real part of the trajectory does not change 
appreciably, we obtain a good agreement for the widths as shown by the 
dashed line in Fig.~\ref{fig1}(b).   

\begin{figure}
\insertplot{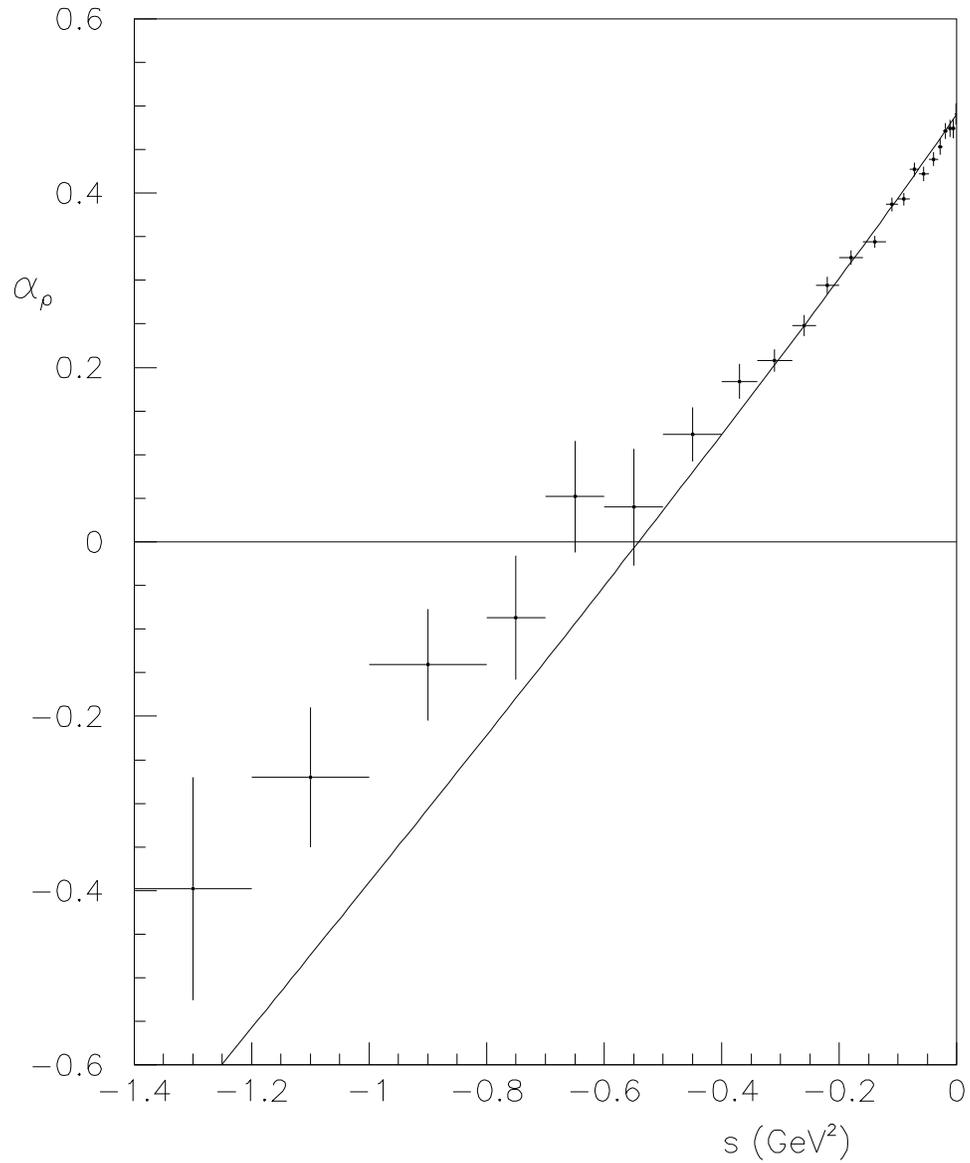}
\vspace{-1cm}
\caption{
Full line is the degenerate $\rho-a_2$ trajectory for negative values of $s$, 
in the scattering region. Points with error bars show the effective 
trajectory obtained by fitting pion charge-exchange scattering 
data~\cite{BARN}.
}
\label{fig3}
\end{figure}

The study of the $f$-trajectory becomes, in this approach, important 
under many aspects. Analyses of total cross sections depend crucially on 
its intercept. Moreover, EXD must be badly broken since attempts to 
describe together $f$ and $\omega$ in this model failed, supporting recent 
findings~\cite{DGMP,M}. The EXD violation can be understood by noticing 
that the width of the $\omega(782)$ is only four percent of the width of 
the $f(1270)$, and it will be difficult to find an analytic function 
satisfying these constraints. Starting from the resonances $f_2(1270), 
\,f_4(2050)$ and $f_6(2510)$, 
which needs confirmation~\cite{CASO}, and the zeroth order trajectory
$\alpha_f(s)= 0.697+0.801\cdot s$, with errors quoted in~\cite{DGMP}, 
we get,
with the higher threshold $s_x=(2 m_{f_2})^2\simeq 6.51 $ GeV$^2$, 
a final intercept $\alpha_f
(0)=0.685$ . At the end of 5th cycle, the parameters are $c_1\simeq 0.155, 
c_2\simeq 0.247$ and $c_x=8.561$. Figs.~\ref{fig2}(a, b) show again the 
${\cal R}e\,\alpha(s)$, ${\cal I}m\,\alpha(s)$
and the width, $\Gamma_f(s)$, compared with the experimental 
points. In Fig.~\ref{fig2} triangles symbolize $\omega(782)$ and 
$\omega_3(1670)$: these resonances are not included in the fit, but 
their masses and widths are drawn to show once again, that EXD 
breaking affects more the widths than the real part of the 
trajectories.

The real part of the degenerate $\rho-a_2$ trajectory has been 
evaluated also for negative values of $s$, in the scattering region. 
From Eq.~(\ref{g2}) a quasi-linear trajectory results with a very small 
positive curvature. In Fig.~\ref{fig3} we compare our result with an effective 
trajectory obtained by fitting pion charge-exchange scattering 
data~\cite{BARN}. Only for $s>-0.4$ GeV$^2$ the agreement with the 
experimental points is very good. The curvature of the analytic 
trajectory has the correct sign, but is not sufficient at large $|s|$. A 
possible solution at large negative $s$ has been proposed in 
Ref.~\cite{FJMPP}, where the fit to the existing data~\cite{BARN,BAT,CASO,SERP,SBA} 
of the differential cross section $d\sigma/dt$ for the process $\pi^-\,p 
\to \pi^0\,n$ reproduces correctly the experimental data. 
However, in~\cite{FJMPP} the trajectory has logarithmic asymptotics and 
it would be difficult to construct a completely solvable 
analytic model in this case.
 
\section{Conclusions}           

In this paper we attempt to systematize the light unflavored mesonic 
trajectories by simultaneous fits to the masses of the resonances and 
their widths. A deeper understanding of EXD plays an important role in 
this program. In addition to the known properties of the S-matrix, 
analyticity and unitarity, other assumptions are needed in order to have 
a completely soluble model. The imaginary part of a trajectory depends 
on the presence of one or more branch points, related to corresponding 
thresholds, for positive values of the argument. The first hypothesis is 
that these thresholds are additive. Moreover, analyticity places an 
upper bound on the asymptotic behaviour of a trajectory. The second 
assumption fixes a particular asymptotic form compatible with this 
constraint. These two assumptions, regarding only the imaginary part 
of the trajectory, determine the analytic structure of the model.

The parameters of the model, the position of the thresholds and their 
weight together with the intercept of the trajectory, have been 
determined for the EXD $\rho - a_2$ trajectories and for the $\rho$ and $f$ 
alone, and the EXD violation has been discussed and 
explained in the context of the model.

Due to the asymptotic behaviour of the real part,  
only a finite number of resonances may lie on a given 
trajectory in this model. 
However, for the examples considered here the bound on the resonance 
masses is too high for putting any constraint on a future search
of new mesonic states. For the $f$ trajectory in fact, a higher 
effective threshold will be probably present. In this respect our 
approach differs from the one in Ref.~\cite{KOB}.

The general agreement with the experimental data, in the time-like and 
space-like regions, of this simple analytical model makes us hopeful about 
its capability in predicting properties of mesonic resonances, not yet 
confirmed or discovered.

\section{Acknowledgment}

We thanks Enrico Predazzi for stimulating discussions on the subject of
this paper. One of us (L.L.J.) is grateful to the Dipartimento di Fisica 
dell'Universit\`a della Calabria and to the Istituto Nazionale di Fisica 
Nucleare - Sezione di Padova e Gruppo Collegato di Cosenza for their warm 
hospitality and financial support. The work of L.L.J. was partly supported by 
INTAS, grant 97-1696, and CRDF, grant UP1-2119.

\vfill\eject

\end{document}